\documentclass[twocolumn,preprintnumbers,superscriptaddress,amsmath,amssymb]{revtex4-1}
\usepackage{float}
\usepackage{graphicx}% Include figure files
\usepackage{dcolumn}% Align table columns on decimal point
\usepackage{bm}% bold math
\usepackage{epstopdf}
\usepackage{xcolor}
\usepackage{textcomp}
\usepackage{soul}% \st
\usepackage{csquotes}
\usepackage{amsbsy}
\usepackage{mathrsfs} % script-like, curvy letters.
\usepackage{amsmath}
\usepackage{bigints} % Big Integration
\usepackage{amsthm,amssymb} % Theorem Proof
\usepackage[utf8]{inputenc}
\usepackage[english]{babel}
\usepackage[shortlabels]{enumitem}
\usepackage{float}
\usepackage{mathrsfs,bigints,mathtools,dsfont}
\usepackage[colorlinks=true,linkcolor=blue,citecolor=blue]{hyperref}%
\usepackage[toc]{appendix}
\usepackage{longtable}

\begin{document}	
\title{{Desynchrony induced by higher-order interactions in triplex metapopulations}}
\author{Palash Kumar Pal}\affiliation{Physics and Applied Mathematics Unit, Indian Statistical Institute, 203 B. T. Road, Kolkata 700108, India}
\author{Md Sayeed Anwar}\affiliation{Physics and Applied Mathematics Unit, Indian Statistical Institute, 203 B. T. Road, Kolkata 700108, India}
\author{Dibakar Ghosh}\email{dibakar@isical.ac.in}\affiliation{Physics and Applied Mathematics Unit, Indian Statistical Institute, 203 B. T. Road, Kolkata 700108, India}	

\begin{abstract}
	In a predator-prey metapopulation, the two traits are adversely related: synchronization and persistence. A decrease in synchrony apparently leads to an increase in persistence and, therefore, necessitates the study of desynchrony in a metapopulation. In this article, we study predator-prey patches that communicate with one another while being interconnected through distinct dispersal structures in the layers of a three-layer multiplex network. We investigate the synchronization phenomenon among the patches of the outer layers by introducing higher-order interactions (specifically three-body interactions) in the middle layer. We observe a decrease in the synchronous behavior or, alternatively, an increase in desynchrony due to the inclusion of group interactions among the patches of the middle layer. The advancement of desynchrony becomes more prominent with increasing strength and numbers of three-way interactions in the middle layer. We analytically validated our numerical results by performing the stability analysis of the referred synchronous solution using the master stability function approach. Additionally, we verify our findings by taking into account two distinct predator-prey models and dispersal topologies, which ultimately assert that the findings are generalizable across various models and dispersal structures.    
\end{abstract}	
		
\maketitle
\section{Introduction}\label{sec1}   
The study of predator-prey systems has received a great deal of attention in the field of ecology because of their importance in different complicated biological processes \cite{berryman1992orgins,taylor1990metapopulations}. In ecology, metapopulation dynamics \cite{chave2004scale} describes the behavior of an ensemble of geographically dispersed populations of identical species that have some degree of mutual dependence. The metapopulation hypothesis highlights the significance of connectedness among the populations since multiple populations working together instead of an isolated population can aid in the survivability of a group of species \cite{holyoak1996persistence,holyoak1996role}. In light of this, it is essential to address geographically distributed population dynamics through the lens of network theory \cite{boccaletti2006complex,wang2003complex}. This interpretation makes sense if the nodes of the network stand in for ecological patches and the connections among them represent migration routes. However, due to the increasing intricacy of the ecological systems, including geographical separation, risky environments, and lack or over-availability of resources, species from one spatial location can interact with other species from different spatial locations in various ways throughout the year \cite{moore1991temporal,schoenly1991temporal,winemiller1996factors} or, the movement of species from a patch can be affected by the simultaneous presence of other species from different patches \cite{koen2007process,bairey2016high,abrudan2015socially}. For these reasons, the framework of metapopulation schematized by classical networks has been generalized in various ways, including multilayer networks that can handle a wide variety of interactions via numerous layers \cite{kivela2014multilayer,boccaletti2014structure}, and higher-order networks (simplicial complex, or hypergraphs) that take into consideration simultaneous interactions comprising three or even more nodes \cite{hypergraph1,beyond_pairwise,majhi2022dynamics}. These higher-order structures have been proven to produce novel features in various dynamical processes \cite{alvarez2021evolutionary,kumar2021evolution}. Multilayer \cite{pilosof2017multilayer,kundu2021persistence} and higher-order \cite{billick1994higher, koen2007process,bairey2016high, abrudan2015socially} networks thus provide the necessary structure for expressing the increasing ecological complexities. In our present study, we will be analyzing a specific type of multilayer design called multiplex networks \cite{battiston2017new,gomez2013diffusion}, where identical sets of interacting nodes are replicated across the layers while maintaining one-to-one correlation among the counterpart nodes that are responsible for the interconnection across the layers.      
\par Synchronization \cite{rosenblum2003synchronization,osipov2007synchronization,anwar2023synchronization}, wherein the system individuals evolve in unison, is one of the fascinating phenomena observed in multiplex networks, which have captivated a lot of attention in the area of network research. There are numerous types of synchronization phenomena that occur in multiplex networks, such as cluster synchronization \cite{della2020symmetries}, antiphase synchronization \cite{chowdhury2021antiphase}, explosive synchronization \cite{zhang2015explosive}, chimera states \cite{majhi2016chimera,majhi2017chimera}, intralayer synchronization \cite{gambuzza2015intra,intra2,anwar2022stability}, interlayer synchronization \cite{sevilla2016inter,leyva2017inter,anwar2022intralayer}, and relay interlayer synchronization \cite{rakshit2021relay,anwar2021relay,drauschke2020effect,leyva2018relay}. This synchronization phenomenon also has widespread interest in ecological populations, where the species display identical fluctuations in time \cite{liebhold2004spatial,holland2008strong,ranta1995synchrony,heino1997synchronous}. In ecology, specifically in predator-prey metapopulations, the synchronous phenomenon is of particular importance since it has been observed and theoretically predicted that synchronization and persistence are negatively correlated with each other in metapopulations \cite{holyoak2000habitat,goldwyn2009small}. There would not be any migrants to act as rescuers if all synchronous species suddenly and catastrophically declined at the same time. In other words, if the patches are all in a synchronized state and a disturbance occurs in one of them, it quickly spreads throughout the entire population, whereas in a desynchronized population, the disturbance is contained within a smaller number of patches. This necessitates the study of desynchronization or, alternatively, abatement of synchronization in predator-prey populations, as it apparently leads to an increasing persistence. Motivated by this, here, we perform a theoretical study to find a way to hinder synchrony on the predator-prey population network arranged in a multiplex framework, where species are subjected to higher-order interactions. The consideration of higher-order interactions in ecological populations is not new \cite{abrams1983arguments,case1981testing,billick1994higher}; however, their role in the reduction of synchronization in multiplex metapopulation structure is yet to be investigated.             
\begin{figure}[ht]
	\includegraphics[scale=0.43]{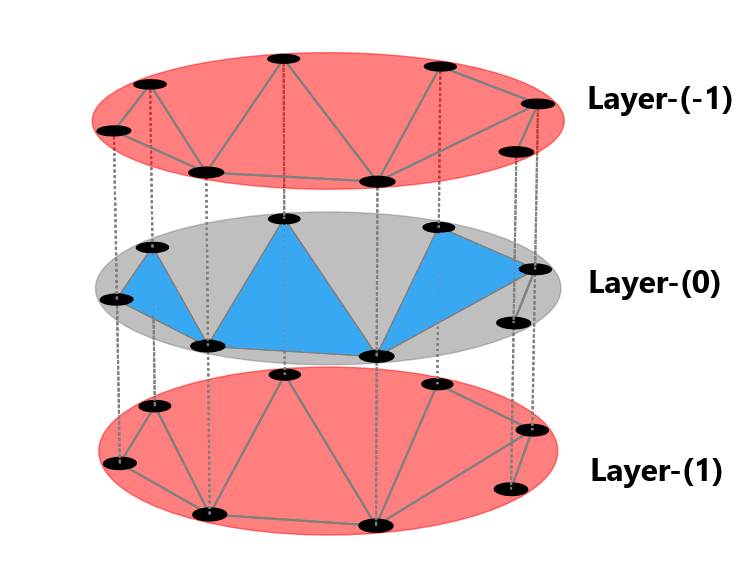}
	\caption{{\bf Schematic diagram of our proposed triplex network architecture.} Each layer is composed of 8 patches, schematized by solid black circles. The layer-(0) in the middle (colored in grey) acts as a relay layer in which the triangles shaded in blue indicate three-way interactions between the patches, and solid black lines represent the pairwise connections between the patches. The outer two layers (layer-(1) and layer-(-1), colored in red) are subjected to only pairwise interactions depicted by solid black lines. The dashed black lines between two adjacent layers indicate the pathway for species migration from one layer to another.}
	\label{fig1}
\end{figure}
\par In this article, we consider a three-layered framework where the layers stand for different geographical regions containing several patches of the same predator-prey species. The connectivity patterns among the patches in the outer layers are subjected to only pairwise diffusive interactions, i.e., species movement from one patch to another depends only on the patches that are connected through pairwise links. Whereas the patches within the middle layer interact with one another through both pairwise and three-way interactions, i.e., species movement from one patch to another depends not only on the patches that are connected through pairwise links but also on the patches connected through $2$-simplices (triangles). Again, the species within a layer may be motivated to go to adjacent geographical locations (i.e., adjacent layers) based on resource availability, which is exemplified by interlayer migration. To reduce the intricacy, we assume species from any patch in one layer can migrate only to its replica-positioned patches in the adjacent layers. In this three-layer multiplex (triplex) framework, we scrutinize the synchronous behavior among the patches in the outer layers, called relay interlayer synchronization \cite{anwar2021relay,rakshit2021relay,leyva2018relay}. Our analysis shows a decrease in synchrony among patches of the outer layers, and as a result, a sufficiently wider region of desynchronization is achieved when we introduce three-way interactions among the patches in the middle layer. We verify our findings analytically by executing the stability analysis of the synchronous solution in terms of the master stability function approach for two different three-dimensional chaotic predator-prey populations.                 
\par The remaining parts of this article are structured as follows. In Section \ref{sec2}, we develop the mathematical framework for predator-prey metapopulation organized on a three-layer multiplex network. Section \ref{Results} elaborately details the results for two different three-species chaotic food chain models. Finally, we sum up all our results and conclude in section \ref{conclusion}.
\section{Framework for triplex metapopulation}\label{sec2}

 We concentrate on the scenario of a multiplex network with three layers (schematized in Fig. \ref{fig1}), each of which contains $N$ number of predator-prey patches amalgamated with one another through dispersal topology. Every patch in a layer is interconnected to its replica patches in adjacent layers. In this way, all the patches from the middle layer are interconnected to their replicas in the two outer layers, and those in the outer layers are intertwined only with their twins in the middle layer. Consequently, the middle layer acts as a relay layer to the indirectly connected outer layers. The pairwise connections among the patches within and between the layers signify species' movement across the patches through diffusive coupling. Furthermore, we assume that the species movement from a patch in the middle layer is not only dependent on its adjacent patches connected through pairwise links but also on the patches that are connected through three-way interactions. In other words, the nodes (patches) in the middle layer are interconnected with each other via both two and three-body interactions simultaneously. However, the patches within the outer layers are interconnected only through pairwise links. Consideration of such three-body interactions, along with pairwise ones, is very much relevant in predator-prey metapopulation dynamics. For instance, a predator can change its targeted prey because of the availability of other prey \cite{koen2007process}. In this example, the two preys do not engage with one another directly; rather, they participate in a three-body interaction, which would be overlooked if just pairwise interactions are considered. Taking all these aspects together, the dynamics of the three-layered metapopulation is given by the following sets of differential equations,          
\begin{subequations}\label{gen_model}
  \begin{multline}
\dot{\bf X}_{-1,i} = f({\bf X}_{-1,i})+\epsilon_{1} \sum\limits_{j=1}^{N}\mathscr{A}^{[-1]}_{ij} G^{(1)}[{\bf X}_{-1,j}-{\bf X}_{-1,i}] 
		\\ + \eta H[{\bf X}_{0,i}-{\bf X}_{-1,i}],
  \end{multline}
  \begin{multline}
    \dot{\bf X}_{0,i} = f({\bf X}_{0,i})+\epsilon_{1}\sum\limits_{j=1}^{{N}}\mathscr{A}^{[0]}_{ij}G^{(1)}[{\bf X}_{0,j}-{\bf X}_{0,i}] 
        \\ +\epsilon_{2} \sum\limits_{j,k=1}^{{N}}\mathscr{A}^{[h]}_{ijk}G^{(2)}({\bf X}_{0,i},{\bf X}_{0,j},{\bf X}_{0,k}) \\ +\eta H[{\bf X}_{1,i}+{\bf X}_{-1,i}- 2{\bf X}_{0,i}],  
  \end{multline}
		\begin{multline}
		\dot{\bf X}_{1,i} = f({\bf X}_{1,i})+\epsilon_{1} \sum\limits_{j=1}^{N}\mathscr{A}^{[1]}_{ij} G^{(1)}[{\bf X}_{1,j}-{\bf X}_{1,i}] 
		\\ + \eta H[{\bf X}_{0,i}-{\bf X}_{1,i}], ~~ i=1,2,\cdots,N    
\end{multline}  
\end{subequations}

where ${\bf X}_{k, i}$ indicates the $d$-dimensional state vector of the $i$-th patch in layer-$k$ with $k=-1,1$ representing the outer layers and $k=0$ corresponds to the middle layer. The intrinsic dynamics of each patch are given by $f({\bf X}_{k, i})$. The real-valued parameters $\epsilon_{1}$ and  $\epsilon_{2}$ represent the strength of pairwise and three-way interactions within the layers, respectively, while $\eta$ indicates the migration strength of populations across the layers. $G^{(1)}: \mathbb{R}^{(d)} \rightarrow  \mathbb{R}^d$ and $H: \mathbb{R}^{(d)} \rightarrow  \mathbb{R}^d$ are the inner-coupling matrices associated with the pairwise interactions within and across the layers, respectively, indicating which species will migrate from one patch to another. The second term in the evolution equation of all the layers indicates the pairwise diffusive interaction between different patches of a particular layer. $\mathscr{A}^{[k]}$ $(k=-1,0,1)$ is the adjacency matrix of layer-$k$ whose entries are defined as $\mathscr{A}^{[k]}_{ij}=1$ if movement of species can be observed between $i$-th and $j$-th patch through the pairwise links while $\mathscr{A}^{[k]}_{ij}=0$ if no such movement is observed. The third term in the dynamics of the middle layer defines the three-way interactions between the species of the patches characterized by the adjacency tensor $\mathscr{A}^{[h]}$ and the interaction function $G^{(2)}({\bf X}_{0,j},{\bf X}_{0,k},{\bf X}_{0, i})$, which can be either linear or nonlinear. The entries of the associated adjacency $\mathscr{A}^{[h]}$ are defined as $\mathscr{A}^{[h]}_{ijk}=1$ if a three-way interactions between the patches $i$, $j$ and $k$ occur, and  $\mathscr{A}^{[h]}_{ijk}=0$ otherwise. Finally, the last term in the dynamics of all the layers represents the migration of species from one layer to its adjacent layers by means of diffusive coupling. The outer layers are symmetric with respect to the middle layers, i.e., the adjacencies $\mathscr{A}^{[-1]}$ and $\mathscr{A}^{[1]}$ have the same structure. In addition to this, we assume that similar to the pairwise interactions, the three-way interactions are also characterized by a diffusive-like coupling function. Therefore, when the interaction function $G^{(2)}$ is linear it takes the form $G^{(2)}({\bf X}_{0,j},{\bf X}_{0,k},{\bf X}_{0, i})= H^{(2)}[{\bf X}_{0,j}+{\bf X}_{0,k}-2{\bf X}_{0,i}]$, where $H^{(2)}: \mathbb{R}^{(d)} \rightarrow  \mathbb{R}^{(d)}$ is the corresponding inner coupling matrix. These three-way interactions thus characterize the dispersal of species between three different patches simultaneously, governed by a linear diffusion scheme. Saying differently, the movement of species from one patch is dependent on more than one patch concurrently. On the other hand, when the three-way interactions are nonlinear diffusive, then there exists a function $g^{(2)}:\mathbb{R}^{(2d)} \rightarrow  \mathbb{R}^{(d)}$ such that $G^{(2)}$ can be represented as 
\begin{align}
    G^{(2)}({\bf X}_{0,j},{\bf X}_{0,k},{\bf X}_{0, i})= g^{(2)} ({\bf X}_{0,j}, {\bf X}_{0,k}) - g^{(2)} ({\bf X}_{0,i}, {\bf X}_{0,i}).
\end{align}
Intuitively, one may interpret this as a generalization of the typical diffusion process on metapopulations, which tends to homogenize the regional variations and as a result, disappear in the situation of equal system states (i.e., in synchronized state). Consequently, in the context of nonlinear three-way interactions, the higher-order coupling function adopts specific forms. For instance, the second-order coupling function could be expressed as $g^{(2)} ({\bf X}_{0,j}, {\bf X}_{0,k})={\bf X}_{0,j}{\bf X}_{0,k}$ and $g^{(2)} ({\bf X}_{0,j}, {\bf X}_{0,k})={\bf X}_{0,j}^{(2)}{\bf X}_{0,k}$, and is referred to as quadratic and cubic diffusion, respectively \cite{muolo2023turing}. Let us observe that the nonlinear diffusion process has been studied in ecology, but mostly when the system dynamics are governed by PDEs \cite{kadota2006positive,zhou2008analysis,sun2014influence}. Here, we extend the study of nonlinear diffusion processes on the networked systems governed by ODEs. Throughout the main text, we specifically consider the cubic diffusion process to represent the three-way interactions. The results with linear and quadratic diffusive couplings are discussed in Appendix \ref{different_couplings}.    
\par We further presume that the dispersal among the patches in the three layers happens randomly with a probability $p$, i.e., species are moving from one patch to another with the probability $p$. Here, we consider $N=100$ patches in each layer interacting with one another through random dispersal topology having probability $p=0.1$. In this way, species from each patch can move randomly to approximately $(N-1)p$ numbers of other patches within a particular layer. Using the network theory concept, one can generate this type of dispersal topology using the Erd\H{o}s-R\'enyi random network algorithm \cite{erdos1959random}. Now, the three-way interactions between the species of the patches in the middle layer are generated by considering all the $2$-simplices (triangles) formed during the creation of random dispersal topology. Our premise for constructing this three-way interaction structure is not exclusive. One can choose any other way to generate these interactions; for example, the species from different patches interact with each other only through three-body connections but not through direct pairwise connections, or there may be a probability to generate a three-way interaction between the species of any three patches. We here consider the simplest way to construct the topology for three-way interactions, that is by considering the $2$-simplices in the random dispersal topology. In other words, here we are assuming that if three patches $i$, $j$, and $k$ are connected with each other in such a way that both $j$ and $k$ are neighbors of $i$, and also $j$ and $k$ are neighbors of each other, then in addition to the pairwise dispersal process among them, there also exists a nonlinear (linear) diffusion process between them. We contemplate the significance of these three-way interactions as a representation of the fact that the migration of species between patches is influenced by more than just pairwise connections. In fact, the movement could rely on multiple neighboring patches concurrently, introducing a layer of complexity beyond linear relationships. 
\section{Results} \label{Results}
 Throughout this section, we investigate a particular synchronization phenomenon called Relay interlayer synchronization (RIS) in our proposed triplex metapopulation framework \eqref{gen_model}. Relay synchronization indicates the synchrony between the patches of two layers indirectly connected via a middle (relay) layer. More precisely, relay synchrony occurs when the replica patches in the outer layers display identical oscillations over time, i.e., ${\bf X}_{1,j}(t)={\bf X}_{-1,j}(t)$ $(j=1,2,\cdots, N)$. It should be noted that relay synchronization does not necessitate that the patches within each layer oscillate in unison, nor that the patches between the outer and middle layer exhibit identical fluctuation over time \cite{leyva2018relay}. This implies that the relay synchronous solution can be achieved even if ${\bf X}_{k,j}(t)\neq {\bf X}_{k, i}(t)$ $(k=-1,0,1; ~ i,j=1,2,\cdots, N)$ and ${\bf X}_{k,j}(t)\neq {\bf X}_{0, j}(t)$ $(k=1,-1)$. In this regard, the invariance of the relay synchronous solution is guaranteed in our considered multiplex metapopulation \eqref{gen_model} due to the choice of diffusive dispersal coupling between the replica patches of the adjacent layers \cite{rakshit2021relay}.
 \par To quantify the relay synchronization state, we introduce the instantaneous synchronization error $E_{RIS}(t)=\dfrac{1}{N}\sum\limits_{j=1}^{N} \|\mathbf{x}_{1,j}(t)-\mathbf{x}_{-1,j}(t)\|$, which is zero when the replica patches of the outer layers fluctuate in unison and non-zero finite when the oscillation of the replica patches in the outer layers is asynchronous. In the following, we will typically consider the time average of the synchronization error to better estimate the transition between the synchronous and asynchronous states. The multiplex network \eqref{gen_model} is therefore evaluated for a period of $3\times10^5$ time steps with integration strep size $\delta t=0.01$ and the last $10^{5}$ time units are taken for calculating the average synchronization error. To better understand the impact of higher-order interactions, specifically three-way interactions, on the emergence of synchronous oscillation among the patches in outer layers, we assume the dispersal topology of each layer to be identical so that the three-way interactions will only play the role of a difference-maker in our investigation. 
 \par In the following subsections, we scrutinize the relay interlayer synchronization behavior by taking into account two distinct three-species chaotic food chain models as the intrinsic dynamics of individual patches. We employ chaotic models as dynamical units since this choice can be instructive for studying synchronization in a more broad scenario where species do not inevitably converge to an equilibrium state \cite{hp}.
 \subsection{Hastings-Powell model}\label{hp}
We start our investigation by considering the Hastings-Powell three-species chaotic food chain model \cite{hp} with one prey, one predator, and one super-predator species in every patch. The interaction among these three species occurs through a Holling type-II functional response. Here, the intrinsic evolution of the patches is given by   
\begin{equation}
	f({\bf X})= \begin{pmatrix*}
    f_{1}(x,y,z) \\\\
    f_{2}(x,y,z) \\\\
    f_{2}(x,y,z)
\end{pmatrix*} = 
\begin{pmatrix*} 
 x(1-x)- \frac{a_{1}xy}{1+b_{1}x} \\\\
 \frac{a_{1}xy}{1+b_{1}x}-\frac{a_{2}yz}{1+b_{2}y}-d_{1}y \\\\
 \frac{a_{2}yz}{1+b_{2}y}-d_{2}z
 \end{pmatrix*},
\end{equation}
Where $x$ corresponds to the species at the lowest food chain level, $y$ is associated with the species that preys upon $x$, and $z$ accounts for the species that preys upon $y$. The system parameter values are kept fixed at $a_1=5.0$, $b_1=3.0$, $a_2=0.1$, $b_2=2.0$, $d_1=0.4$, $d_2=0.01$ so that each patch exhibits chaotic dynamics when uncoupled. We assume that only the prey species move from one patch to another within each layer, i.e., all the entries except the first entry of the inner-coupling matrix $G^{(1)}$ are zero. More precisely, only the prey species take part in pairwise interactions among the patches within each layer. Similarly, only the prey species take part in the three-way interactions within the middle layer. Thus, we consider the three-way interaction function $G^{(2)}$, characterized by cubic diffusion as $G^{(2)}({\bf X}_{0, i},{\bf X}_{0, j},{\bf X}_{0, k})=[x^{2}_{0,j}x_{0,k}-x^{3}_{0, i},0,0]$, i.e., the three-way interaction between the prey species of any three patches $i$, $j$, and $k$ is not realized by linear diffusive coupling. Rather, we choose a nonlinear diffusive coupling form. This particular coupling form, for instance, represents the combined effect of prey species from patch $j$ and $k$ on the prey species of patch $i$, with patch $j$ being more effective than patch $k$. However, we assume that all three species can migrate from one layer to its adjacent layer via pairwise interlayer links, i.e., all the entries of the coupling matrix $H$ are zero except the diagonal ones. Here, we consider that within the layers, only one species (in our case, prey species) can move from one patch to another, but all three species can move from one layer to another layer. Intuitively, this can be considered as the scenario that under normal circumstances, one or more species moves within local regions (i.e., between the patches of a particular layer), and when the circumstances are extreme, like natural hazards or risky environment, then all the species migrate from their location to an adequately farthest location (i.e., migration to different layers). Nonetheless, our findings remain robust even when introducing alternative scenarios. This encompasses situations where all three species are permitted to traverse both within and between the layers, as well as cases where either one or two species are granted this mobility within and across layers. The results with some of these alternative scenarios are discussed in Appendix \ref{different_couplings}.
\begin{figure}[ht]
	\centerline{
	   \includegraphics[scale=0.25]{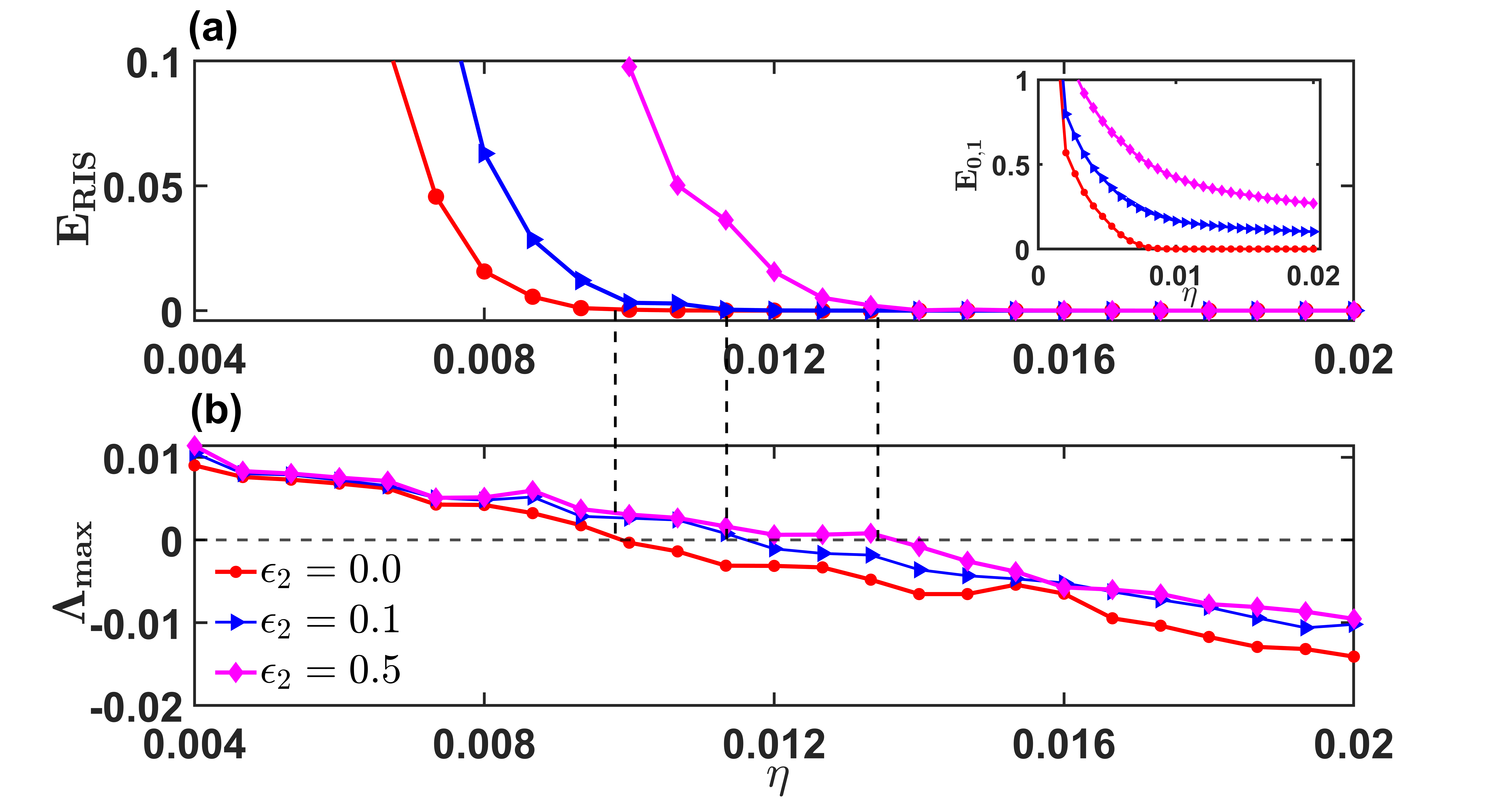}}
   \caption{{\bf Synchronization between the patches of the outer layers of the multiplex metapopulation with HP model in each patch.} (a) Synchronization error $E_{RIS}$ as a function of interlayer migration strength $\eta$ for three different values of three-way interaction strength: $\epsilon_{2}=0$ (red curve), $\epsilon_{2}=0.1$ (blue curve), and $\epsilon_{2}=0.5$ (magenta curve). The inset depicts the synchronization error between the middle $(k=0)$ and the outer $(k=1)$ layer defined as $E_{0,1}(t)=\dfrac{1}{N}\sum\limits_{j=1}^{N}\|\mathbf{x}_{0,j}(t)-\mathbf{x}_{1,j}(t)\|$. (b) The maximum Lyapunov exponent $\Lambda_{max}$ evaluated from the variational Eq. \eqref{transverse_hp} (in Appendix \ref{stability}) as a function of $\eta$ for the same set of values of $\epsilon_{2}$ as in (a). The pairwise intralayer dispersal strength is kept fixed at $\epsilon_{1}=0.2$ in all the subplots. }
   \label{fig2}
\end{figure}
\par In order to study the emergence of relay interlayer synchronization phenomena in the multiplex framework \eqref{gen_model}, we start by evaluating the synchronization error $E_{RIS}$ by varying the strength of migration $\eta$ between the layers, for different three-way interaction strength $\epsilon_{2}$ and a fixed value of intralayer dispersal strengths $\epsilon_{1}=0.2$. The corresponding results are depicted in Fig. \ref{fig2}(a). The intralayer coupling strengths $\epsilon_{1}$ and $\epsilon_{2}$ are taken in such a way that the patches within a particular layer do not display any identical fluctuations over time (synchrony). For $\epsilon_{2}=0$, i.e., when there is no three-way interaction between the species of the patches in the middle layer, the replica patches of the indirectly connected layers (outer layers) achieve a synchronous state for a critical value of interlayer migration strength $\eta \approx 0.01$ (red curve in Fig. \ref{fig2}(a)). In addition to that, the middle (relay) layer also displays identical fluctuation with the outer layers as shown in the red curve in the inset of Fig. \ref{fig2}(a) where the synchronization error $E_{0,1}$ between the middle and the outer layer becomes zero at approximately same critical interlayer migration strength. These two synchronous behaviors are obvious when there are no three-way interactions in the middle layer, as in this particular scenario, all three layers are identical. Now to investigate the impact of three-way interactions in the emergence of a relay synchronous state, we gradually increase the value of $\epsilon_{2}$. For $\epsilon_{2}=0.1$ (shown in the blue curve), one can perceive that a comparably larger critical interlayer migration strength $\eta$ is needed for the achievement of relay synchronous state. The occurrence of the synchronous solution is delayed further with the increasing three-way coupling strength. For $\epsilon_{2}=0.5$, a synchronous oscillation between the patches of outer layers is achieved at $\eta \approx 0.014$. Moreover, we observe that with the introduction of three-way interactions between the patches of the middle layer, the patches in the middle (relay) layer start displaying asynchronous fluctuation with that of the outer layers (shown in the inset where $E_{0,1}>0$). {\it Therefore, the inclusion of higher-order interactions in the relay layer of a triplex metapopulation delays the occurrence of synchronous oscillation between the patches of outer layers when compared with the triplex metapopulation having only pairwise interactions among the patches. In other words, the group interactions between the patches of the middle layer induce desynchronization between the outer layers.}
\par To validate the acquired result, we proceed through the stability of the relay synchronous solution using the master stability function approach \cite{pecora1998master}. Evaluating the maximum Lyapunov exponent $\Lambda_{max}$ transverse to the synchronous manifold ${\bf X}_{1, i}={\bf X}_{-1, i}$ gives the necessary condition for stable synchronous oscillation (detailed in Appendix \ref{stability}). The negative value of $\Lambda_{max}$ with varying coupling strengths indicates its stability. In Fig. \ref{fig2}(b), we plot the transverse maximum Lyapunov exponent $\Lambda_{max}$ as a function of interlayer migration strength $\eta$ for the same set of intralayer coupling strengths $\epsilon_{1}$ and $\epsilon_{2}$ as in Fig. \ref{fig2}(a). One can observe that the $\Lambda_{max}$ curves become negative at the same critical values of $\eta$ for which the value of synchronization error $E_{RIS}$ is zero, indicated by the dashed vertical lines. Thus, our observation regarding the delay in the occurrence of synchrony among the patches of outer layers with the introduction of three-way interactions in the relay layer is validated analytically using the master stability approach.
\begin{figure}[ht]
	\centerline{
		\includegraphics[scale=0.25]{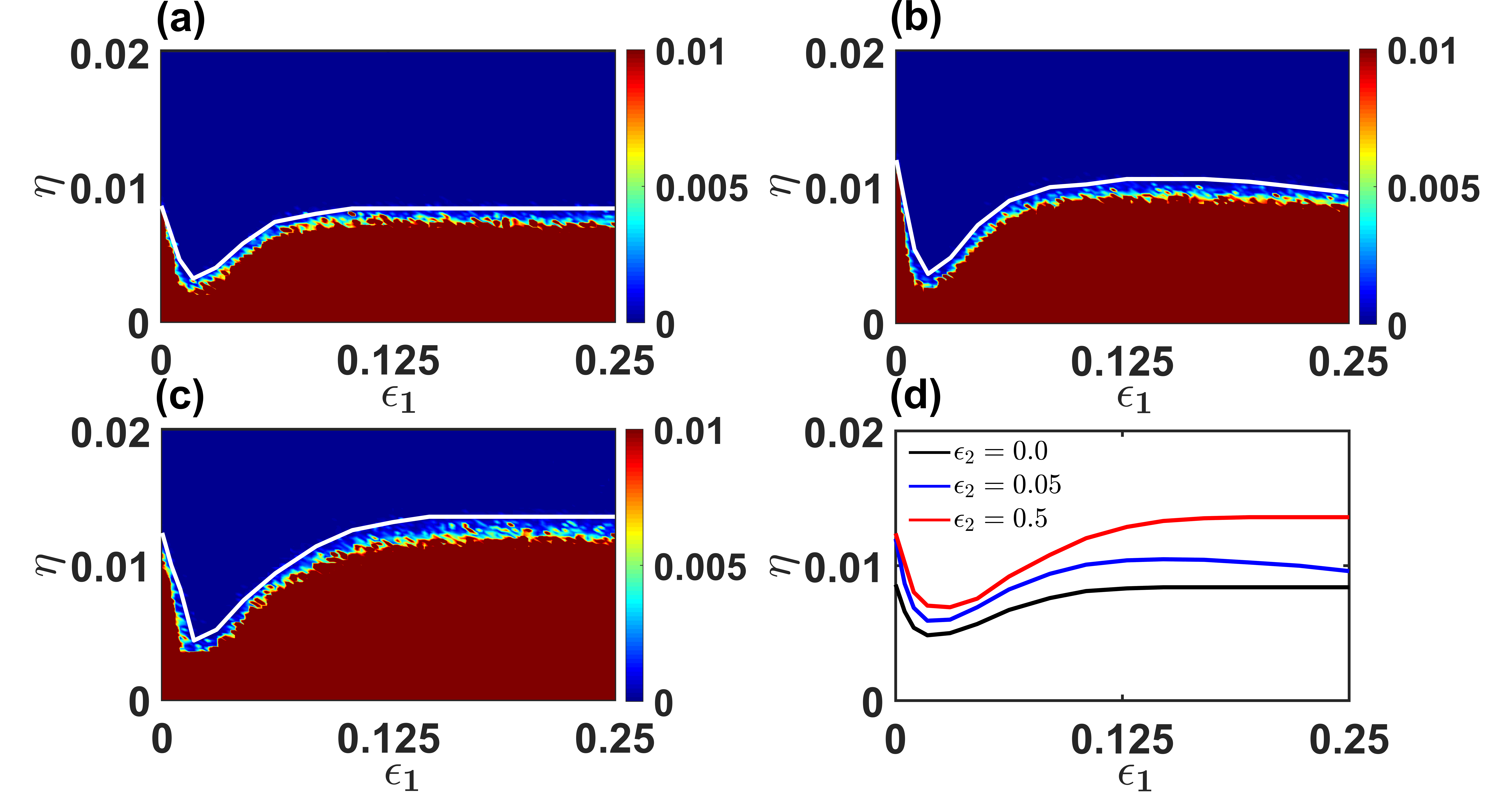}}
	\caption{ {\bf Region of synchronization and desynchronization in ($\epsilon_1,\eta$) parameter plane for the three values of three-body interaction coupling strength}: (a) $\epsilon_{2}=0.0$, (b) $\epsilon_{2}=0.05$, (c) $\epsilon_{2}=0.5$. The color bar represents the variation of synchronization error $E_{RIS}$. The solid white curves immersed on each parameter space denote the boundary which divides the region of synchrony and desynchrony and is obtained by solving the variational equation \eqref{transverse_hp} for $\Lambda_{max}(\epsilon_{1},\eta)=0.0$. (d) The boundary curves corresponding to the three values of $\epsilon_{2}$ (see legend) are plotted separately, which asserts the increase in desynchrony region with increasing $\epsilon_{2}$.}
	\label{fig3}
\end{figure}
\par Thereafter, to scrutinize the complete scenario of relay synchronization in a wider range of parameter values, we evaluate the synchronization error $E_{RIS}$ by simultaneously varying the interlayer migration strength $\eta$ and pairwise dispersal strength $\epsilon_{1}$ for different values of three-way interaction strength $\epsilon_{2}$. Figure \ref{fig3} delineates the corresponding results. The solid white lines in each subfigure define the boundary between the synchronous (blue region) and asynchronous (red region) oscillation, obtained by solving the transverse error dynamics for the calculation of maximum Lyapunov exponent $\Lambda_{max}$. In particular, these curves correspond to $\Lambda_{max}(\epsilon_{1},\eta)=0$. While investigating the relay synchronization phenomenon for different values of $\epsilon_{2}$, it can be observed from Fig. \ref{fig3}(a) that the triplex network with solely pairwise connections $(\epsilon_{2}=0$) exhibits relay synchronous behavior even in the absence of any pairwise interaction $\epsilon_{1}$. The critical value of $\eta$ for the emergence of synchronous oscillation decreases with increasing value of $\epsilon_{1}$ and attains the lowest value $\eta \approx 0.0032$ at $\epsilon_{1} \approx 0.0175$. Beyond that, further increment of $\epsilon_{1}$ shifts the threshold for achieving synchrony again toward higher values of $\eta$ until the intralayer dispersal strength $\epsilon_{1}$ is reached at $\epsilon_{1} \approx 0.105$. If we increase the value of $\epsilon_{1}$ further, the replica patches in outer layers display identical fluctuation for $\eta \gtrapprox 0.0085$ independent of the value of $\epsilon_{1}$. A similar scenario is observed in Fig. \ref{fig3}(b), where we introduce three-way interactions in the middle layer with coupling strength $\epsilon_{2}=0.05$. However, in this case, we perceive that the critical threshold to achieve the synchronous fluctuation is relatively higher than the solely pairwise situation. This results in a reduced synchronization region and an enhanced desynchrony region. This behavior becomes more prominent when we increase the three-way coupling strength to $\epsilon_{2}=0.5$, as depicted in Fig. \ref{fig3}(c). {\it Therefore, rigorously plotting the region of synchrony and desynchrony, we can admirably differentiate the relay synchronous behavior for the cases of pairwise and three-way interactions, which reasserts our remark that the three-way interaction in relay layer delays the occurrence of synchrony between the replica patches of the outer layers, or saying differently induces desynchrony among those patches.} 
\begin{figure}[ht]
	\centerline{
		\includegraphics[scale=0.23]{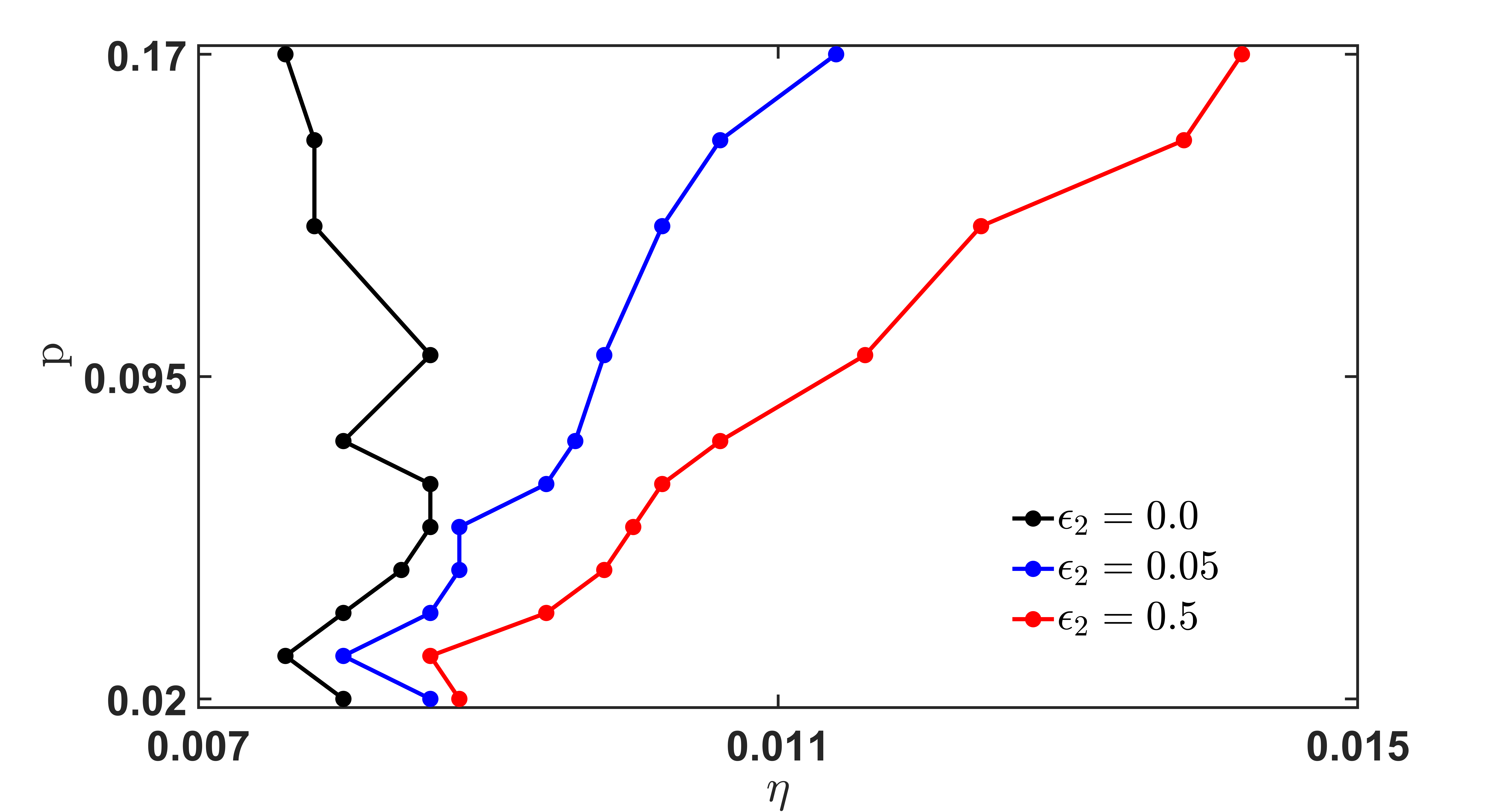}}
	\caption{{\bf Critical interlayer migration strength $\eta$ with varying probability $p$ for generating random dispersal topology.} Three curves represent the critical values of $\eta$ where the synchronous state occurs for three different values of three-way interaction strength $\epsilon_{2}$: black curve corresponds to the thresholds for $\epsilon_{2}=0.0$, while blue, and red account for $\epsilon_{2}=0.05$, and $\epsilon_{2}=0.5$, respectively. The regions on the left of these critical curves are the regions of asynchronous oscillation, while the regions beyond these curves correspond to the regions of synchronous oscillation. The figure reflects that with increasing $p$ and $\epsilon_{2}$, the synchronous oscillation emerges for relatively larger values of $\eta$, i.e., an increase in desynchrony is observed. The intralayer pairwise coupling strength is fixed at $\epsilon_{1}=0.2$ for the calculation of all the critical points.}
	\label{fig4}
\end{figure}
\par Until now, the results we have discussed are associated with fixed random probability $p$. But, $p$ is one of the most important parameters in our study as it controls the species' movement from one patch to another. When $p=0$, no species movement occurs among the patches. The species from one patch can move to more patches with increasing $p$, and when $p=1$, the species can move from one patch to every other patch. Consequently, with increasing $p$, the number of three-way interactions also increases in the middle layer. Therefore, to elucidate the effect of probability $p$ on the occurrence of relay synchrony, we calculate the critical value of interlayer migration strength $\eta$ with varying probability $p$. The corresponding result is portrayed in Fig. \ref{fig4}. The black curve corresponds to the completely pairwise scenario ($\epsilon_{2}=0$), while the blue and red ones represent the critical curve for $\epsilon_{2}=0.05$ and $\epsilon_{2}=0.5$, where the three-way interactions are present in the relay layer. The domains on the left of these critical curves are the domains of asynchronous oscillation, while the domains beyond these curves (i.e., on the right) correspond to the region of synchronous oscillation. As observed, for $\epsilon_{2}=0$, the critical coupling for the emergence of synchronous oscillation is more or less similar with increasing $p$, i.e., when there is no three-way diffusion in the middle layer, due to the presence of only pairwise diffusion the occurrence of synchronous oscillation between the patches of the outer layers is not greatly affected by the number of pairwise diffusions, rather we can observe a slight enhancement on the occurrence of synchronous oscillation with increasing connection probability in the regime of large $p$. However, when the three-body diffusion process comes into play (i.e., $\epsilon_{2}>0$), an interesting phenomenon can be observed. For a minute fraction of three-way interaction strength ($\epsilon_{2}=0.05$), we observe that with the increasing value of $p$, i.e., with the increasing number of three-body diffusions, the synchronous oscillation emerges at relatively higher values of interlayer migration strength $\eta$. This eventually leads to a larger region of asynchronous oscillation as compared to the scenario where three-body diffusions are absent. The emergence of synchronous oscillation is further delayed with the increment in the three-way coupling strength value $\epsilon_{2}$, as delineated for $\epsilon_{2}=0.5$. {\it Therefore, higher-order interactions serve as crucial contributors to the attenuation of synchronous oscillations, playing a pivotal role in shaping the outcome. Particularly, the higher number of three-way interactions and larger value of three-way coupling strength together results in a larger desynchronization region and subsequently delays the emergence of synchronous oscillation between the patches of outer layers up to a large value of interlayer migration strength.}
\subsection{ Gakkhar-Naji model} \label{GN_model}
To provide further evidence that the results obtained due to the introduction of higher-order interaction in the relay layer of the triplex metapopulation are not model dependent, here, we conduct the analysis while taking into account another three-species predator-prey model proposed by Gakkhar and Naji \cite{sgp}. Here, the interaction among the species occurs through a Holling type-II functional response, and the corresponding evolution equations are given by 
\begin{equation}
\begin{array}{l}
	f({\bf X})= 
\begin{pmatrix*} 
{x}\big[1-{x}-\frac{{y}}{1+{w}_1{x}}-\frac{{z}}{1+{w}_2{x}+{w}_3{y}}\big]\\\\
 {y}\big[\frac{{w}_4{x}}{1+{w}_1{x}}-{w}_5-\frac{{w}_6{z}}{1+{w}_2{x}+{w}_3{y}}\big] \\\\
 {z}\big[\frac{{w}_7{x}+{w}_8{y}}{1+{w}_2{x}+{w}_3{y}}-{w}_9\big]
 \end{pmatrix*},
 \end{array}
\end{equation}
where $x$ corresponds to the species at the lowest food chain level, $y$ is associated with the species that preys upon $x$, and $z$ accounts for the species that prey upon both $x$ and $y$. The system parameter values are kept fixed at ${w}_1=1.44,{w}_2=5.0,{w}_3=8.0,{w}_4=1.0,{w}_5=0.1,{w}_6=0.1,{w}_7=0.1,{w}_8=0.1,{w}_9=0.01$, so that each patch exhibits chaotic dynamics when uncoupled.

\begin{figure}[ht]
 	\centerline{
 		\includegraphics[scale=0.25]{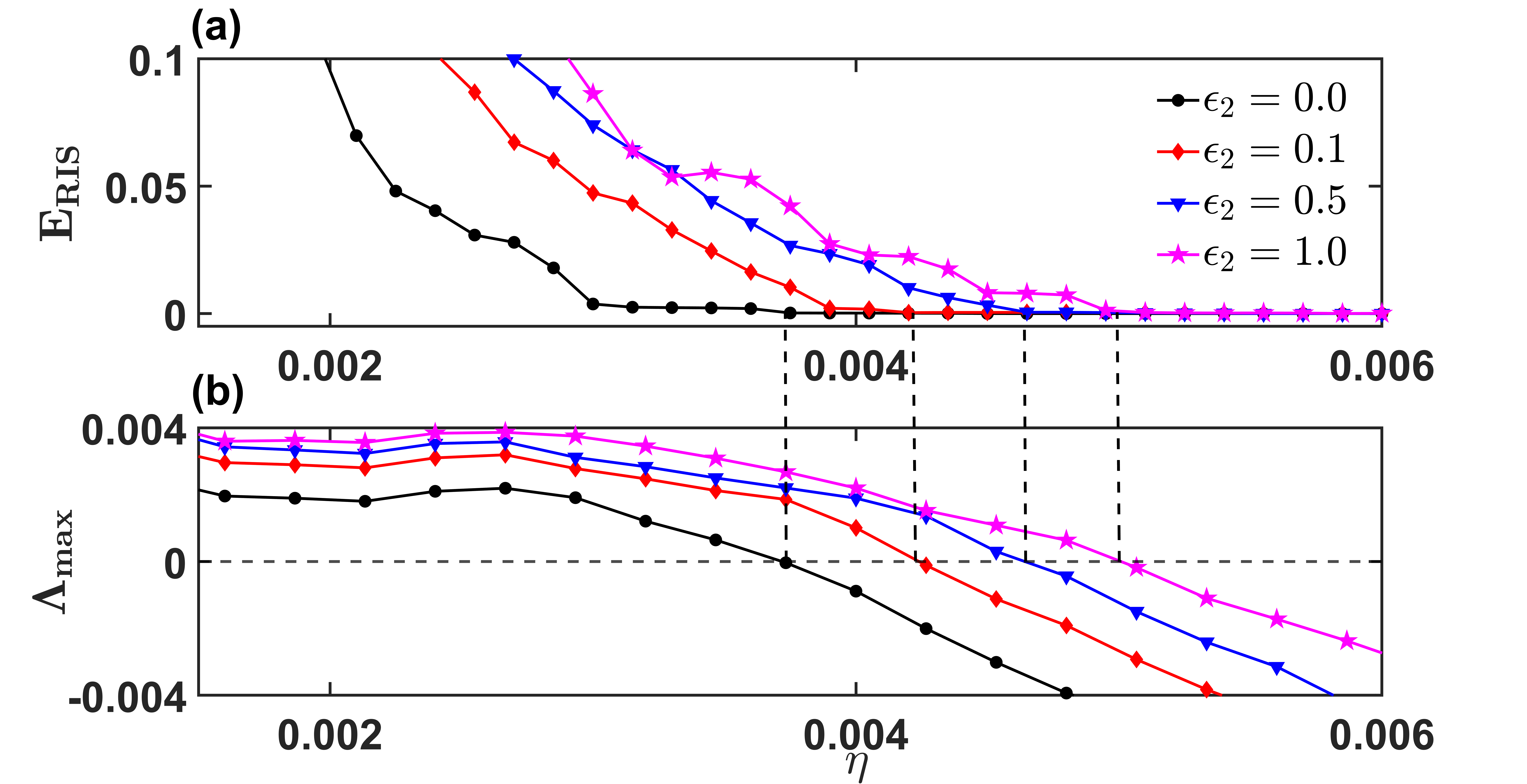}}
 	\caption{{\bf Synchronization between the patches of the outer layers of the multiplex metapopulation with the Gakkhar-Naji model in each patch.} (a) Synchronization error $E_{RIS}$ as a function of interlayer migration strength $\eta$ for four distinct values of three-way interaction strength: $\epsilon_{2}=0.0$ (black curve), $\epsilon_{2}=1.0$ (red curve), $\epsilon_{2}=1.0$ (blue curve), and $\epsilon_{2}=1.0$ (magenta curve). (b) Maximum Lyapunov exponent $\Lambda_{max}$ evaluated from the variational Eq. \eqref{variation_eq1} as a function of $\eta$ for the same set of values of $\epsilon_{2}$ as in (a). The pairwise intralayer dispersal strength is kept fixed at $\epsilon_{1}=0.012$ in all the subplots.}
 	\label{fig5}
 \end{figure}

We evaluate the synchronization error $E_{RIS}$ and plot them as a function of interlayer migration strength $\eta$ in Fig. \ref{fig5}(a) for four different values of three-way interaction strength $\epsilon_{2}$ keeping the intralayer pairwise interaction strength fixed at $\epsilon_{1}=0.012$. For $\epsilon_{2}=0.0$, i.e., when the triplex metapopulation is subjected to only pairwise interactions, the replica patches of two outer layers start oscillating identically ($E_{RIS}=0$) beyond $\eta \approx 0.0037$, displayed by the black curve. On the other hand, for $\epsilon_{2}=0.1,0.5,1.0$, i.e., when the middle layer is subjected to both pairwise and three-way interactions, the relay synchronization emerges beyond relatively higher interlayer migration strengths $\eta \approx 0.0037,0.0042,0.0048$ successively, which are depicted in the red, blue, and magenta curves, respectively. The obtained results are accompanied by plotting the curves of maximum Lyapunov exponent $\Lambda_{max}$ in Fig. \ref{fig5}(b), where the curve of $\Lambda_{max}$ crosses the zero-line and becomes negative exactly at the same critical coupling $\eta$ for which $E_{RIS}$ is zero, as indicated by the vertical dashed lines.
\begin{figure}[ht]
	\centerline{
		\includegraphics[scale=0.25]{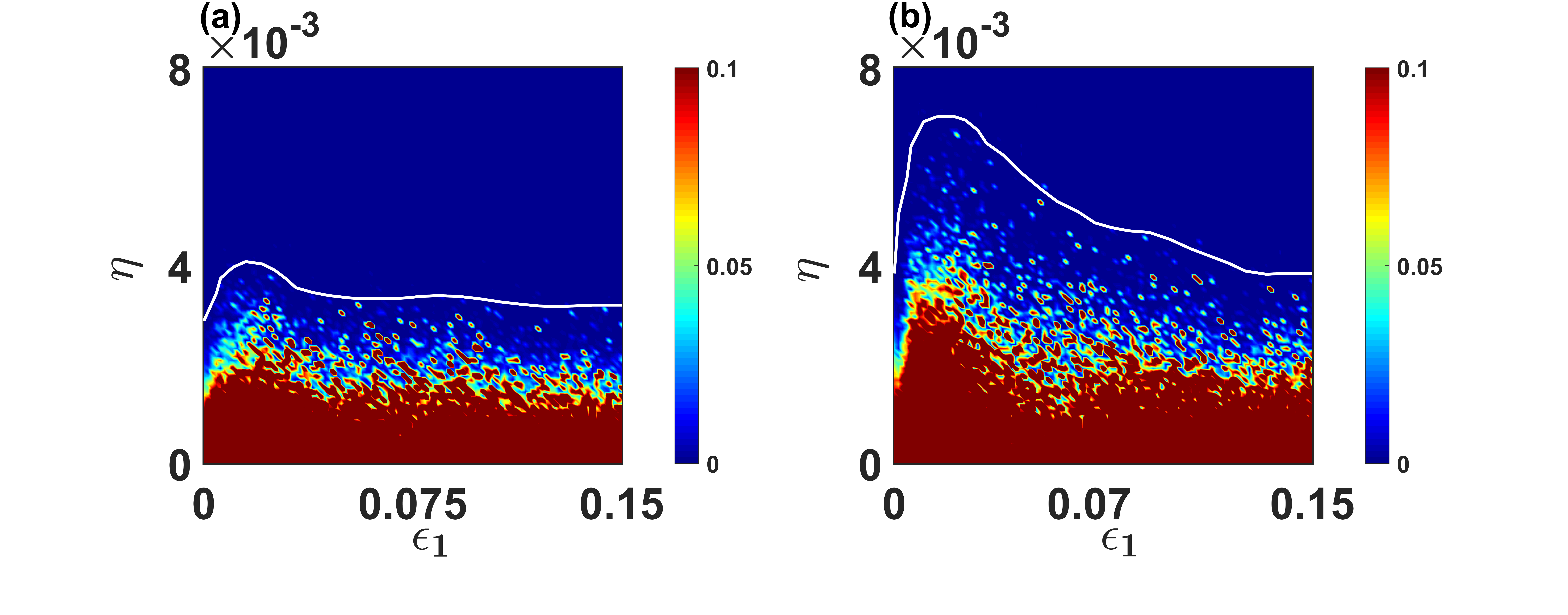}}
	\caption{  {\bf Region of synchronization and desynchronization in ($\epsilon_1,\eta$) parameter plane for two distinct values of three-body interaction coupling strength}: (a) $\epsilon_{2}=0.0$, (b) $\epsilon_{2}=1.0$. The color bar represents the variation of synchronization error $E_{RIS}$. The solid white curves superimposed on each parameter space denote the boundary that divides the region of synchrony and desynchrony and is obtained by solving the variational equation \eqref{variation_eq1} for $\Lambda_{max}(\epsilon_{1},\eta)=0.0$.}
	\label{fig6}
\end{figure}
In addition to that, we also evaluate the region of synchrony and desynchrony by simultaneously varying the intralayer pairwise coupling strength $\epsilon_{1}$ and interlayer migration strength $\eta$ for two different values of three-way interaction strength $\epsilon_{2}=0$ (Fig. \ref{fig6}(a)) and $\epsilon_{2}=1.0$ (Fig. \ref{fig7}(b)). One can observe that the region of synchrony (desynchrony) decreases (increases) adequately as we introduce the three-way interaction in the middle layer. {\it Therefore, from the above results, we can conclude that our observed phenomenon regarding the diminishment of synchronous fluctuation between the patches of outer layers due to the introduction of three-way interactions in the middle layer is not specific to a particular model, rather, it follows in general.}  
 \section{Discussion}\label{conclusion}               
Because of its importance to the notion of species migration and persistence, metapopulation dynamics have been the subject of research in a wide range of subdisciplines within ecology. In many cases, the identical fluctuations of species (synchrony) within a large community of organisms hinder the metapopulation persistence, apparently leading to extinction \cite{holyoak2000habitat}. That means if all the patches experienced declines in an enormous amount at the same time, there would not be any migrants who act as rescue agents. This necessitates the study of asynchrony (desynchrony) within the metapopulation dynamics, as it plays a vital role in increasing metapopulation persistence \cite{kundu2017survivability}. Our present study aims to find a way to decrease (increase) synchrony (asynchrony) within the metapopulation dynamics. 
\par To effectively capture ecological complexity, it is necessary to take into account a wide range of related aspects, including dynamical, functional, and structural characteristics. Species in one area may be compelled to go to another depending on factors including geographic isolation, environmental hazards, and resource scarcity, or the movement of species from one patch to another may be affected by a group of species from different patches simultaneously. This forces us to think about two different ways of species interaction: first, the species from various patches interact with each other through many-body interactions, and second, the species interact with one another via different dispersal patterns schematized by multilayer frameworks. We have analyzed the metapopulation model, where species are spread out across a three-layered multiplex patchwork, with predator and prey species interacting with one another. Furthermore, the species in the middle layer are subjected to three-body interactions. Our study reflects that the synchronous fluctuations among patches in the outer layers are delayed by the presence of three-way interactions in the middle layer. The abatement (improvement) of synchrony (desynchrony) is greatly dependent on the strength and number of three-way interactions among the species. The desynchrony becomes more prominent with increasing strength and numbers of three-way interactions in the middle layer. We validated our findings analytically by performing the stability of the synchronous solution using the master stability function approach. Moreover, we demonstrate that our obtained results are generalizable beyond a specific model. We have also verified that apart from random dispersal topology, our findings are valid with other dispersal topologies between the patches of the layers. For example, the result with small-world dispersal topology is illustrated in Appendix \ref{small-world}. We believe that our study can provide a better theoretical understanding of the impact of many-body interactions on the dynamics of ecological systems and pave the way for further research on the persistence of species. A natural extension of our work would be to study the effect of higher-order interactions on the synchronization phenomenon in a generic multilayer ecological framework where the movement of species across various layers of patches would not be confined to a rigid one-to-one correspondence, as typically observed in a multiplex framework.          

\appendix
\section{Stability analysis of relay interlayer synchronization state in triplex metapopulation} \label{stability}
By inspecting the facts that the inter-layer interactions between the patches are subjected to diffusive coupling, and the structure of the outer layers are identical (i.e., $\mathscr{A}^{[-1]}=\mathscr{A}^{[1]}$), one can immediately conclude that the relay synchronous solution ${\bf X}_{1, i}={\bf X}_{-1, i}$ $(i=1,2,\cdots, N)$ is a trivial solution of equation \eqref{gen_model}. Now, the question is whether the solution can become unstable (or, alternatively, maintain its stability) when triggered by a modest perturbation. To investigate this, we introduce small deviations from the synchronous solution, denoted by $\delta{\bf X}_{i}=\delta{\bf X}_{1, i}-\delta{\bf X}_{-1, i}$, and linearize the equation \eqref{gen_model} using Taylor series expansion up to first order. This eventually gives us the variational equation as follows, 
\begin{equation}\label{variation_eq1}
    \begin{array}{l}
    \delta\dot{\bf X}_{i} = [{J}f({\bf X}_{i})-\eta H]{\delta\bf X}_{i}+\epsilon_{1} \sum\limits_{j=1}^{N}\mathscr{A}^{[1]}_{ij} G^{(1)}[{\delta\bf X}_{j}-{\delta\bf X}_{i}],
    \end{array}
\end{equation}
where $Jf$ denotes the Jacobian matrix of $f$ and ${\bf X}_{i}$ signifies the synchronous solution ${\bf X}_{1, i}={\bf X}_{-1, i}$ that satisfies
\begin{equation}
\begin{array}{l}
		\dot{\bf X}_{i} = f({\bf X}_{i})+\epsilon_{1} \sum\limits_{j=1}^{N}\mathscr{A}^{[1]}_{ij} G^{(1)}[{\bf X}_{j}-{\bf X}_{i}] + \eta H[{\bf X}_{0,i}-{\bf X}_{i}].
  \end{array}
  \end{equation}
In the preceding equation, ${\bf X}_{0, i}$ represents the state variable of the $i$-th patch in the middle layer while the outer layers are synchronized with one another and satisfy the following evolution equation 
  \begin{equation}
      \begin{array}{l}
		\dot{\bf X}_{0,i} = f({\bf X}_{0,i})+\epsilon_{1}\sum\limits_{j=1}^{{N}}\mathscr{A}^{[0]}_{ij}G^{(1)}[{\bf X}_{0,j}-{\bf X}_{0,i}] 
        \\\\~~~~~~~ +\epsilon_{2} \sum\limits_{j,k=1}^{{N}}\mathscr{A}^{[h]}_{ijk}G^{(2)}({\bf X}_{0,i},{\bf X}_{0,j},{\bf X}_{0,k}) +2\eta H[{\bf X}_{i}-{\bf X}_{0,i}].
  \end{array}
\end{equation}
The stake variables ${\delta\bf X}_{i}$ in the variational equation \eqref{variation_eq1} evolve transverse to the referenced synchronous solution, and therefore, to have a stable synchronization state, it is necessary that these transverse modes must die out in time. Hence, evaluating the maximum Lyapunov exponent $\Lambda_{max}$ transverse to the synchronous solution as a function of system parameters will give the necessary condition for stability. Wherever $\Lambda_{max}$ becomes negative, perturbations transverse to synchronous solution go to extinction, and as a result, the referenced solution becomes stable.  
\par For illustration, the variational equation of the HP model with our chosen coupling schemes is as follows,
\begin{equation} \label{transverse_hp}
\begin{array}{l}
\delta\dot{{x}}_i=(1-2{x}_{i}-\frac{a_1{ y}_{i}}{(1+b_1{ x}_{i})^2})\delta{ x}_i-\frac{a_1{ x}_{i}}{1+b_1{x}_{i}}\delta{y}_i\\\\\;\;\;\;\;\;\;\;\;\;\;\;\;\;\;\;\;\;\;\;\;\;\;\;\;\;\;\;\;\;\;\;\;\;+\epsilon_1\sum_{j=1}^{N}\mathscr{A}_{i,j}^{[1]}(\delta{x}_j-\delta{x}_i)-\eta\delta{x}_i, 
\\\\\
\delta\dot{{y}}_i=\frac{a_1{y}_{i}}{(1+b_1{x}_{i})^2}\delta{x}_i+(\frac{a_1{x}_{i}}{(1+b_1{x}_{i})}-\frac{a_2{\bf z}_{1,i}}{(1+b_2{\bf y}_{1,i})^2}-d_1)\delta{y}_i \\\\\;\;\;\;\;\;\;\;\;\;\;\;\;\;\;\;\;\;\;\;\;\;\;\;\;\;\;\;\;\;\;\;\;\; -\frac{a_2{y}_{i}}{(1+b_2{y}_{i})}\delta{z}_i-\eta\delta{y}_i,
\\\\
\delta\dot{{z}}_i=\frac{a_2{z}_{i}}{(1+b_2{y}_{i})^2}\delta{y}_i+(\frac{a_2{y}_{i}}{(1+b_2{y}_{i})}-d_2)\delta{z}_i-\eta\delta{z}_i,
\end{array}
\end{equation} 
where $({\delta x}_i,{\delta y}_i,{\delta z}_i)=({ x}_{1,i}-{ x}_{-1,i}, \; { y}_{1,i}-{y}_{-1,i},\; { z}_{1,i}-{z}_{-1,i})$, and $({ x}_{i}, {y}_{i}, {z}_{i})$ is the state variable of the synchronization manifold. Solving the above equation for the calculation of maximum Lyapunov exponent $\Lambda_{max}$ as the function of coupling parameters gives the necessary condition for a stable synchronous state. 

\section{Small-world dispersal topology} \label{small-world}
We consider another complex dispersal topology, specifically small-world \cite{watts1998collective} topology, to elucidate that the results we obtained are not limited to only random dispersal structure within the patches in a layer. In the case of animal movement, most of them disperse over relatively short distances (i.e., to adjacent patches), while only a minority migrate over substantially large distances, which resembles the small-world network structure. Here, we consider $N=100$ patches interact with each other through the small-world dispersal connectivity mechanism with average degree $\langle k \rangle=8$ and probability $p_{sw}=0.15$, i.e., species from each patch can disperse in average to its $8$ neighboring patches, and from few patches, species can move to longer distanced patches with probability $p_{sw}=0.15$. We assume that the dynamics of each patch are governed by the Hastings-Powell three-species food chain model as discussed in section \ref{hp} with the same coupling schemes. 
 \begin{figure}[ht]
	\centerline{
		\includegraphics[scale=0.25]{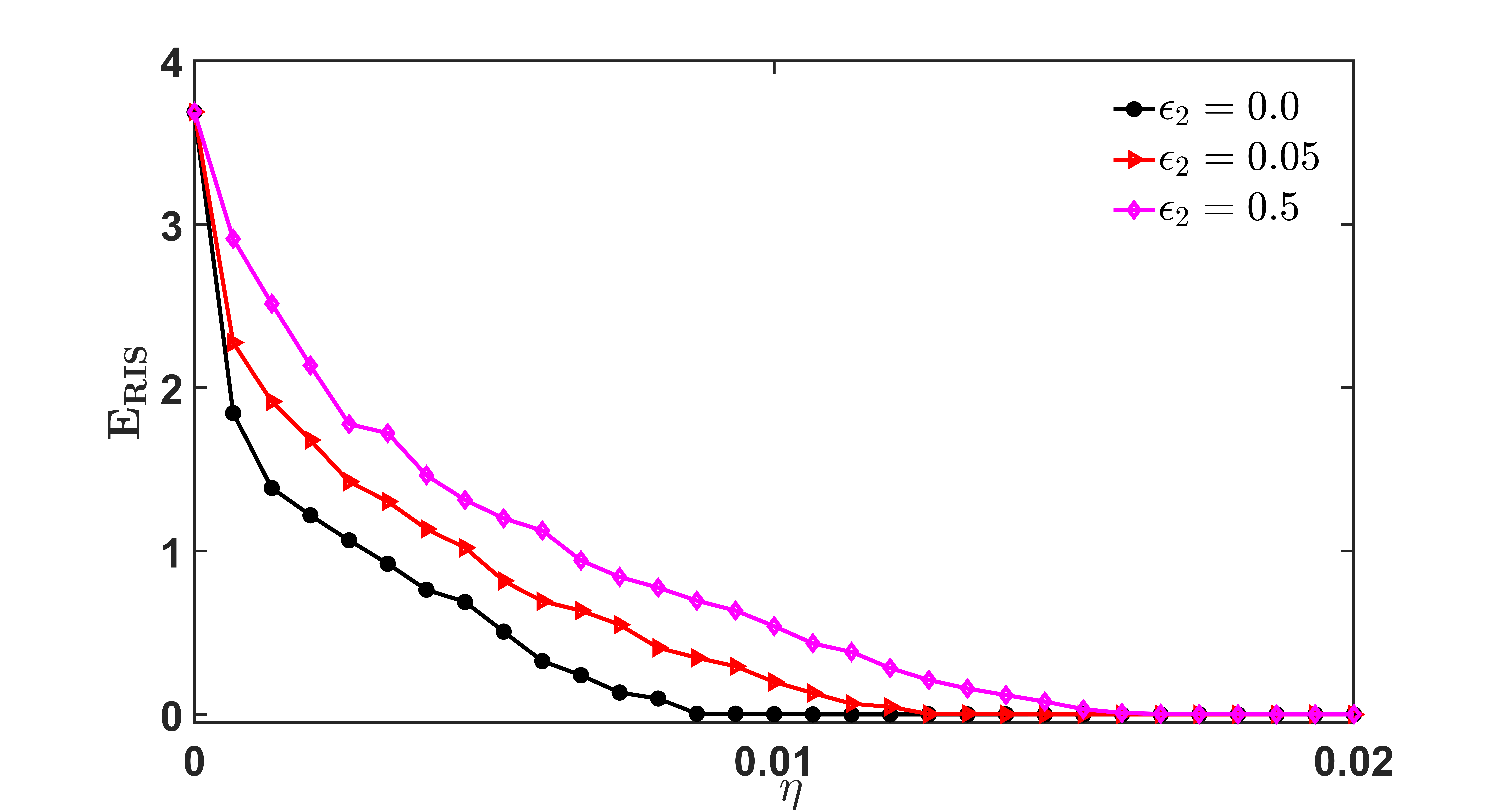}}
	\caption{{\bf Synchronization between the patches of the outer layers of the multiplex metapopulation with small-world dispersal topology in each layer with average degree $\langle k \rangle =8$ and rewiring probability $p_{sw}=0.15$.} Synchronization error $E_{RIS}$ as a function of interlayer migration strength $\eta$ for three distinct values of three-way interaction strength: $\epsilon_{2}=0$ (black curve), $\epsilon_{2}=0.05$ (red curve), and $\epsilon_{2}=0.5$ (magenta curve). The pairwise intralayer dispersal strength is kept fixed at $\epsilon_{1}=0.1$. By increasing the three-way interaction strength $\epsilon_2$, a decrease in synchronization is observed.}
	\label{fig7}
\end{figure} 
\par To investigate the relay synchronization phenomenon (i.e., the synchronous fluctuation among the patches of the outer layers), we plot the synchronization error $E_{RIS}$ as a function of interlayer migration strength $\eta$ for different three-way interaction strength $\epsilon_{2}$ in Fig. \ref{fig7}. For $\epsilon_{2}=0$, i.e., in the absence of three-way interactions in the relay layer, the synchronous fluctuations between the replica patches of the outer layers emerge at $\eta \approx 0.009$. As the three-way interactions are introduced in the middle layer, we observe that the critical strength for achieving the synchrony moves to a relatively larger value. For $\epsilon_{2}=0.05$, synchrony emerges at $\eta \approx 0.013$. The achievement of synchrony is delayed further as we increase the three-way coupling strength to $\epsilon_{2}=0.5$.  
\begin{figure}[ht]
	\centerline{
		\includegraphics[scale=0.25]{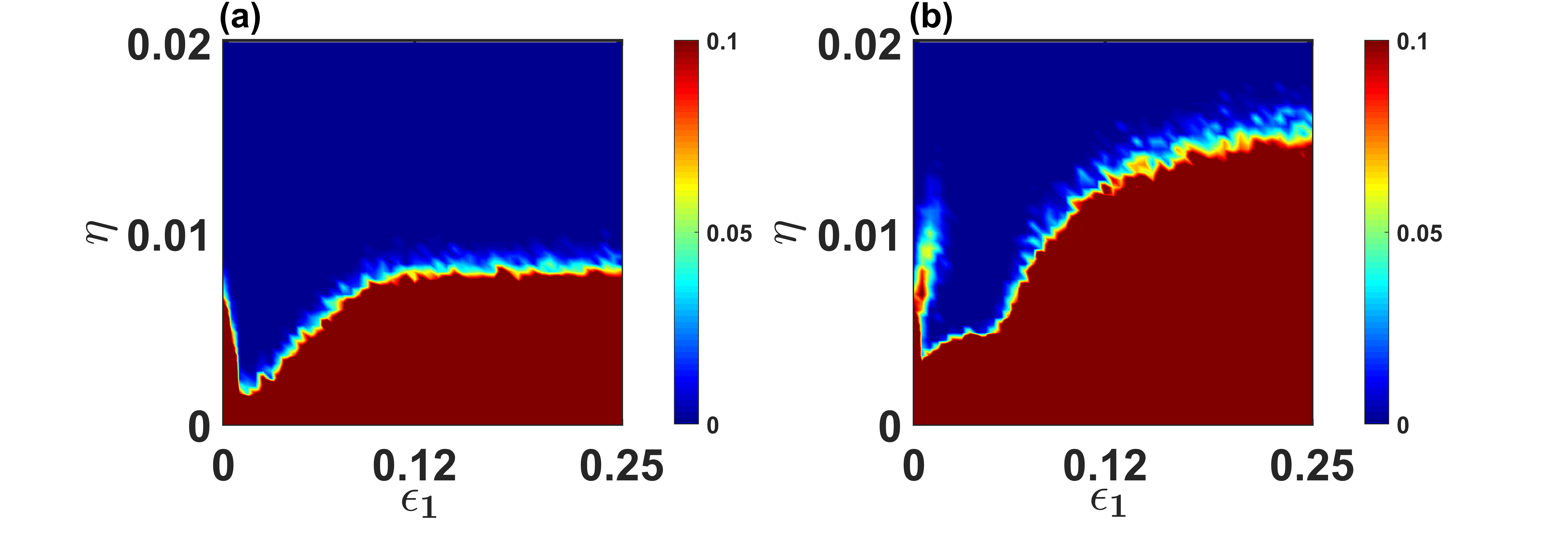}}
	\caption{{\bf Region of synchronization and desynchronization in ($\epsilon_1,\eta$) parameter plane for two distinct values of three-body interaction coupling strength}: (a) $\epsilon_{2}=0.0$, (b) $\epsilon_{2}=0.5$. The color bar represents the variation of synchronization error $E^{RIS}$. A larger region of desynchrony is obtained by the inclusion of higher-order interaction in the relay layer.}
	\label{fig8}
\end{figure}
\par In addition to this, in Fig. \ref{fig8}, we evaluate the synchronization error $E_{RIS}$ by simultaneously varying the coupling strengths $\epsilon_{1}$ and $\eta$ for two different instances of three-way coupling strength: $\epsilon_{2}=0$ (Fig. \ref{fig8}(a)) and $\epsilon_{2}=0.5$ (Fig. \ref{fig8}(b)). As observed, the region of desynchronization (red region) greatly enhances or, alternatively, the synchrony region (blue region) shrinks with the inclusion of three-way interactions. Therefore, our obtained result that the three-way interactions induce desynchronization among the patches of the outer layer holds true even for the small-world dispersal topology within the layers.

%%%%%%%%%%%%%%%%%%%%%%%%%%%%%%%%%%%%%%%%%%%%%%%%%%%%%%%%%%%%%%%%%

\section{Results with other three-way interaction schemes} \label{different_couplings}
To provide further evidence that the results obtained due to the introduction of higher-order interactions in the relay layer of the triplex metapopulation hold in general, we here investigate the outcomes with some other three-way interaction schemes other than the nonlinear (cubic) diffusion interactions, namely linear and quadratic diffusive interactions. In the preceding analysis, we limited intralayer movement to just one species, namely the prey. Expanding upon this, here we explore additional scenarios. This includes instances where all three species possess the ability to navigate within and between layers, along with variations where either one or two species are endowed with the capability for intralayer and interlayer mobility. To investigate all the results we once again choose the individual dynamics of the patches governed by Hastings-Powell's three-species chaotic food chain model.    
\subsection{Linear diffusive coupling scheme}
\begin{figure*}
    \centering
    \includegraphics[scale=0.4]{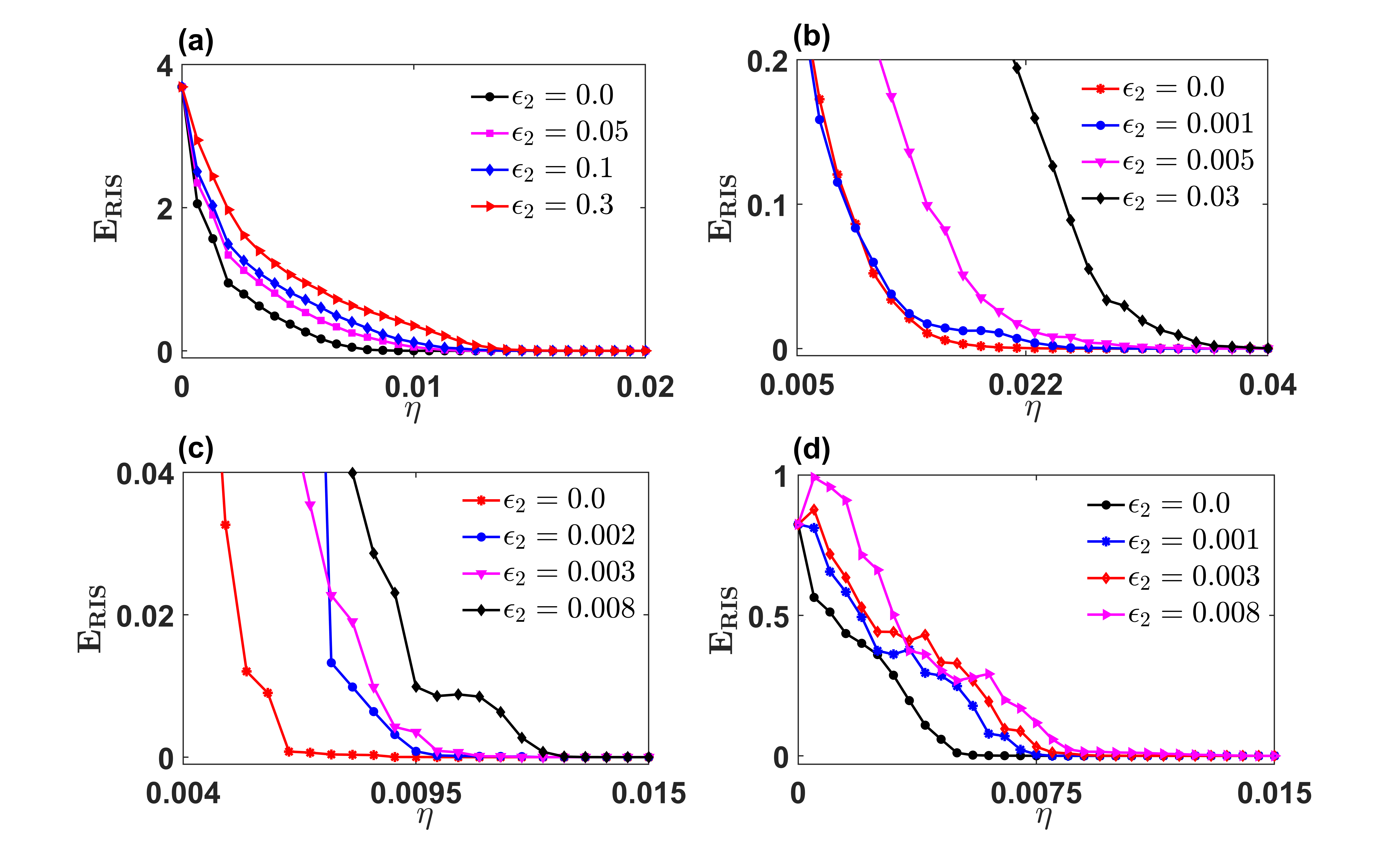}
    \caption{{\bf Synchronization between the patches of the outer layers of the multiplex metapopulation with linear diffusive three-way interactions.} (a) Synchronization error $E_{RIS}$ as a function of interlayer migration strength $\eta$ for four distinct values of three-way interaction strength: $\epsilon_{2}=0$ (black curve), $\epsilon_{2}=0.05$ (magenta curve), $\epsilon_{2}=0.1$ (blue curve), and $\epsilon_{2}=0.3$ (red curve). The pairwise intralayer dispersal strength is kept fixed at $\epsilon_{1}=0.2$. Here only the prey species are allowed to move within the layers and between the layers all three species movement is allowed, (b) $E_{RIS}$ as a function of $\eta$ for $\epsilon_{2}=0,0.001,0.005,0.03$ with $\epsilon_{1}$ fixed at $\epsilon_1=0.001$, when only the predator species are allowed to migrate both within and across the layers, (c) $E_{RIS}$ as a function of $\eta$ for $\epsilon_{2}=0,0.002,0.003,0.008$ with $\epsilon_{1}$ fixed at $\epsilon_1=0.001$, when only the predator and super-predator species are allowed to migrate both within and across the layers, (d) $E_{RIS}$ as a function of $\eta$ for $\epsilon_{2}=0,0.001,0.003,0.008$ with $\epsilon_{1}$ fixed at $\epsilon_1=0.001$, when all three species are allowed to migrate both within and across the layers. In all the cases it can be observed that with the introduction of higher-order interactions, the occurrence of synchronization is delayed.}
    \label{linear_diffusive}
\end{figure*}
Here, we assume that the three-way interactions between the patches of the relay layer are governed by linear diffusive coupling, i.e., $G^{(2)}({\bf X}_{0,j},{\bf X}_{0,k},{\bf X}_{0, i})= H^{(2)}[{\bf X}_{0,j}+{\bf X}_{0,k}-2{\bf X}_{0, i}]$. In this case, we further consider four different instances of species movements within and across the layers. The corresponding results are depicted in Fig.~\ref{linear_diffusive}, where we plot the variation of average relay synchronization error $E_{RIS}$ as a function of interlayer coupling $\eta$ for different values of higher-order coupling strength $\epsilon_{2}$ while keeping the pairwise coupling strength $\epsilon_{1}$ at a nominal value.
\par At first, we consider the species movement the same as before, i.e., only the prey species are allowed to move within the layers, and all three species can traverse across the layers. The only difference is here the three-way interactions in the relay layer are characterized by linear diffusive coupling. In this case, we plot the $E_{RIS}$ by varying $\eta$ for $\epsilon_2=0.0, 0.05, 0.1,\; \mbox{and}\; 0.3$, keeping $\epsilon_{1}$ fixed at $\epsilon_{1}=0.2$ [see Fig.~\ref{linear_diffusive}(a)]. We can observe that with the introduction of higher-order interactions $(\epsilon_{2}=0.05)$, the critical coupling for the achievement of synchronization increases compared to the case of only pairwise interactions $(\epsilon_{2}=0)$. The achievement of synchronization is delayed further as we increase $\epsilon_{2}$ to higher values.
\par In Fig.~\ref{linear_diffusive}(b), we plot the result by considering the situation when only the predator species are allowed to move within and across the layers. Here also with increasing strength of three-body interactions, desynchrony can be induced at those values of $\eta$ for which relay synchrony occurs with only pairwise interactions.
\par A qualitatively similar behavior can be observed when (i) both predator and super-predator species are allowed for intralayer and interlayer migration [see Fig.~\ref{linear_diffusive}(c)] and (ii) all three species are permitted to traverse both within and between the layers [see Fig.~\ref{linear_diffusive}(d)]. {\it Hence, the significant outcome we derived, which showcases how three-way interactions lead to a state of desynchrony among the patches in the outer layers, remains consistent even when altering the three-way interaction structure to linear diffusive interactions. This robustness also holds in the presence of diverse mechanisms governing species movement.}

\subsection{Quadratic diffusion}
\begin{figure}
    \centerline{
    \includegraphics[scale=0.2]{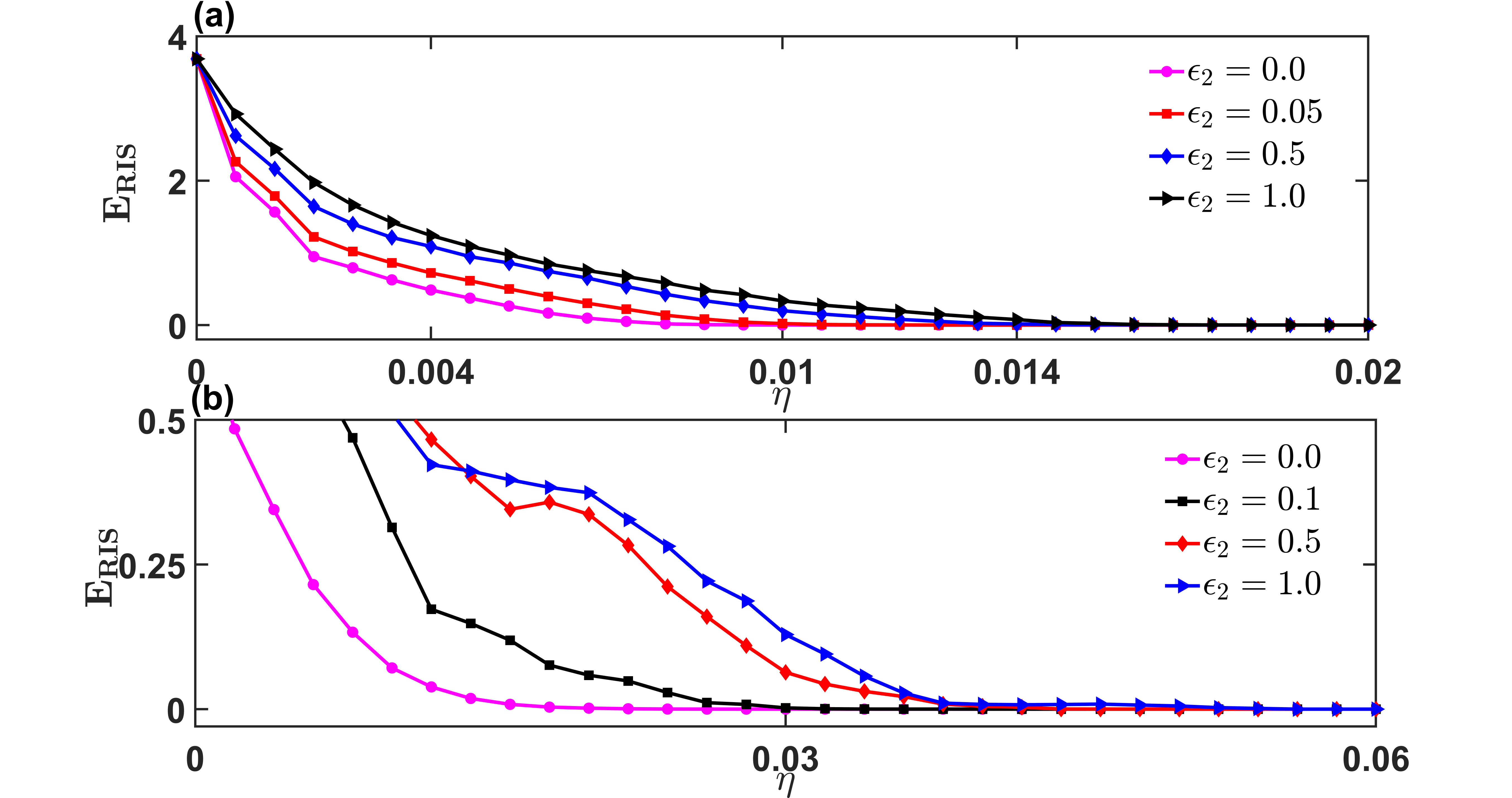}}
    \caption{{\bf Synchronization between the patches of the outer layers of the multiplex metapopulation with quadratic diffusive three-way interactions.}(a) Synchronization error $E_{RIS}$ as a function of interlayer migration strength $\eta$ for four distinct values of three-way interaction strength: $\epsilon_{2}=0$ (magenta curve), $\epsilon_{2}=0.05$ (red curve), $\epsilon_{2}=0.5$ (blue curve), and $\epsilon_{2}=1$ (black curve). The pairwise intralayer dispersal strength is kept fixed at $\epsilon_{1}=0.2$. Here only the prey species are allowed to move within the layers, and between the layers, all three species movement is allowed, (b) $E_{RIS}$ as a function of $\eta$ for $\epsilon_{2}=0,0.1,0.5,1$ with $\epsilon_{1}$ fixed at $\epsilon_1=0.001$, when only the predator species are allowed to migrate both within and across the layers.}
    \label{quadratic_diffusive}
\end{figure}
Here, the three-way interactions are characterized by another nonlinear diffusion scheme, namely quadratic diffusion. We delve into two distinct scenarios of species movement: One is when only the prey species are permitted to move within the layers, and all three species are allowed to traverse across the layers. The other is when only the predator species are allowed to migrate both within and between the layers. The results corresponding to these two instances are delineated in Figs. \ref{quadratic_diffusive}(a) and \ref{quadratic_diffusive}(b), respectively, by plotting $E_{RIS}$ as a function of $\eta$ for various higher-order coupling strength $\epsilon_{2}$. It can be observed that with increasing higher-order coupling strength, the synchronization emerges at relatively higher values of $\eta$. {\it The outcomes thus once again justify our claim that the introduction of the higher-order interactions induces desynchrony among the patches of the outer layers of a triplex metapopulation.}

\bibliographystyle{apsrev4-1} % Tell BibTeX which bibliography style to use
\bibliography{modified}% Produces the bibliography via BibTeX.

\end{document}